\documentclass[12pt]{article}
\usepackage{amsfonts}
\usepackage{bbm}
\usepackage{graphicx}
\usepackage{mathrsfs}
\usepackage{color}
\usepackage{geometry}             
\usepackage{amssymb}
\usepackage{amsmath}
\usepackage{epstopdf}
\usepackage{cases}
\usepackage{cite}

\usepackage[hyperindex=true,
          pdfstartview=FitH,
          bookmarksnumbered=true,
          bookmarksopen=true,
          citecolor=blue,
          linkcolor=blue,
          colorlinks=true,
          pdfborder=001,
          unicode]{hyperref}

\begin{document}

\title{Fluids and vortex from constrained fluctuations around
C-metric black holes}
\author{Xin Hao${}^{1}$\thanks{{\em email}: \href{mailto:shanehowe@mail.nankai.edu.cn}
{shanehowe@mail.nankai.edu.cn}},
Bin Wu${}^{2}$\thanks{{\em email}: \href{mailto:binwu@nwu.edu.cn}
{binwu@nwu.edu.cn}}
and Liu Zhao${}^{1}$\thanks{{\em email}: \href{mailto:lzhao@nankai.edu.cn}
{lzhao@nankai.edu.cn}, correspondence author.}\\
${}^1$ School of Physics, Nankai University, Tianjin 300071, China\\
${}^2$ Department of Physics, Northwest University, Xian 710069, China
}
\date{}
\maketitle

\begin{abstract}
By foliating the four-dimensional C-metric black hole spacetime,
we consider a kind of initial-value-like formulation of the vacuum
Einstein's equation, the holographic initial data is a double consisting
of the induced metric and the Brown-York energy momentum tensor on an
arbitrary initial hypersurface. Then by perturbing the initial
data that generates the background spacetime, it is shown that,
in an appropriate limit, the fluctuation modes are governed by
the continuity equation and the compressible Navier-Stokes equation which
describe the momentum transport in non-relativistic viscous
fluid on a flat Newtonian space. It turns out that the flat space fluid
behaves as a pure vortex and the viscosity to entropy
ratio is subjected to the black hole acceleration.
\end{abstract}
\newpage

\section{Introduction}

In the past two decades the development of fundamental physics
has been greatly promoted by the recognition of the holographic
principle which emerges in the study of black holes. This bold
principle was originally put forward by 't Hooft \cite{t'Hooft}
and Susskind \cite{susskind}, and first realized concretely by
Maldacena \cite{Maldacena:1997re} in the context of string
theory. In this implementation a conjectured equivalence
was established between the
supersymmetric Yang-Mills gauge theory and the superstring
theory. Although the origin
of the connection is still mysterious, this well-known AdS/CFT
correspondence has been applied in many aspects as an efficient
tool to perform analytical calculations in strongly coupled
systems \cite{Polchinski:2001tt,Policastro:2001yc,Son:2002sd}.
A canonical application of this correspondence is the analysis
of shear viscosity in the strongly coupled SYM theory
\cite{Polchinski:2001tt}, in which the hydrodynamics arises as
a classical description for the behavior of any interacting
quantum field theory at long-wavelength and low-frequency. Then
in the framework of AdS/CFT correspondence a connection between
gravity in asymptotically AdS and relativistic hydrodynamics in
one less dimension was revealed
\cite{Policastro:2002se,Bhattacharyya:2008jc,Baier:2007ix,
Haack:2008cp,Bhattacharyya:2008ji,Ashok:2013jda}, and this
is known as the Gravity/Fluid correspondence, and the
relativistic gravitational field equations were reduced
into the incompressible Navier-Stokes equation in a appropriate
scaling limit \cite{Bhattacharyya:2008kq}.

Actually, the Gravity/Fluid correspondence can be realized
independent of AdS/CFT correspondence, and it's origin
could be dated back to the late 1970's, long before the
holographic principle was proposed. Here we refer to the
membrane paradigm \cite{Damor,Price,Damour:2008ji,
Eling:2009sj}, in which the local dynamical quantities
of black holes was first studied ``holographically'', and
shown to be governed by the Damour-Navier-Stokes equations. Despite
the technical differences, the viscosity/entropy ratio from
membrane paradigm at horizon and AdS/CFT at spatial infinity
was surprisingly the same, which indicates a correlation
between these two approaches \cite{Kovtun:2003wp,Son:2007vk,
Iqbal:2008by}. Under such background the Gravity/Fluid
correspondence at arbitrary cutoff was constructed
\cite{Bredberg:2010ky}. In it's original version this was
achieved by considering linearized Einstein equations while
making hydrodynamic expansion and imposing appropriate boundary
conditions. To be specific, the scalar and tensor modes of the
fluctuations are fixed by the ingoing-wave boundary conditions
on the horizon and Dirichlet boundary conditions at the cutoff,
and the only dynamical modes are the vector fluctuations
governed by the linearized Navier-Stokes equations. Shortly
afterwards in \cite{strominger1} the analysis was improved to
the nonlinear case, more importantly the hydrodynamic
expansions and the boundary conditions are shown to be
mathematically equivalent to the near horizon expansions
and Petrov-like boundary conditions respectively
\cite{Strominger2}.

In the recent years this reformulated Gravity/Fluid
correspondence was greatly generalized
\cite{Compere,cai1,Eling,Compere:2012mt,Eling:2011cl,
Bai:2012ci,Zou:2013ix,Banerjee:2012iz,Hu:2013dza,
Niu:2011gu,Nakayama:2011bu,Cai,Cai:2012mg,1204.2029,
xiaoning,Huang:2011kj,Ying2,Wu:2013kqa,Ling:2013kua,WB1,WB2,
Cai:2014ywa,Cai:2014sua}, especially in \cite{HWZ} we studied a
type of constrained perturbations around a class of black holes
with curved horizons, and in the near horizon region (later this
was generalized to finite cutoff \cite{Hao:2015zxa}), we find
that such kind of Petrov-type-fluctuations could be mapped to a
forced compressible viscous fluid in flat space of one less
dimension. In these works we went beyond the framework of the
bulk/boundary duality, and the relaxation of bulk/boundary
restriction may possibly be a motivator and reminder of deeper
understanding about the holographic principle, also we expect
that this further generalization of the Gravity/Fluid correspondence
could unearth more potentiality of gravity as a powerful tool in
the study of fluid dynamics. But we are far from being
optimistic, since we have not yet achieved a complete
construction, one of the major obstacles is the unusual body
force of the dual fluid system pertaining to the surface stress.
To make sense of the external force we need to apply our
analysis to other type of solutions of the Einstein's equation.
So, in this paper we take that first step and study a fluid dual
of a well-known four-dimensional accelerating black hole
solution, i.e. the C-metric black hole. We find that in this
case the form of the external force is better understood, and
more intriguingly, there will be a vortex in the dual fluid
system.

\section{The C-metric reformulated in an appropriate coordinate system}

The whole construction relies on the C-metric solution of the vacuum Einstein
equation \cite{Griffiths:2006tk}. The line element reads
\begin{align}
\mathrm{d}s^2 = \frac{1}{\mathscr{A}} \Big(-Q \mathrm{d}t^2
+ \frac{\mathrm{d}r^2}{Q} + \frac{r^2}{P}\mathrm{d}\theta^2
+ \frac{P r^2 \sin^2 \theta}{(1+\alpha r_h)^2} \mathrm{d}
\varphi^2 \Big),    \label{CM}
\end{align}
where
\begin{align}
&\mathscr{A}(r,\theta)
= (1 + \alpha r \cos \theta)^2, \nonumber\\
&Q(r) = (1 - \alpha^2 r^2)\bigg(1 - \frac{r_h}{r}\bigg),  \nonumber\\
&P(\theta) = 1 + \alpha r_h \cos \theta,  \nonumber
\end{align}
and $0 \leq \alpha r_h < 1$. This solution can be viewed as a one-parameter
generalization of the Schwarzschild metric and is interpreted as an accelerating
black hole solution, with the black hole event horizon located at $r=r_h$, and the
acceleration horizon at $r=\frac{1}{\alpha}$. The positive parameter $\alpha$ corresponds to the proper acceleration of the black hole.

To construct the hydrodynamic equations in isotropic
coordinates, it is desirable to introduce a conformal isotropic coordinate
system in the angular part of the line element. The new coordinates $(x^1, x^2)$
replace the old ones $(\theta,\varphi)$ via the relations
\begin{align*}
& x^1 = w(\theta) \cos \varphi,                     \\
& x^2 = w(\theta) \sin \varphi,
\end{align*}
where
\begin{align*}
w(\theta)
= \Bigg( \Big(\frac{P}{\sin \theta}\Big)^{ \alpha r_h}
\tan \frac{\theta}{2}\Bigg)^{\frac{1}{1 -  \alpha r_h}}
\end{align*}
plays the role of a radial coordinate on the $(w,\varphi)$ ``plane'' and
it is a univariate function in $\theta$ ranging from 0 to $+\infty$.
Though seemingly weird, the construction of this new coordinate system is
straightforward.
Using the above transformation,
the line element \eqref{CM} could be rewritten under the coordinates
$x^\mu = (t,r,x^1,x^2)$ in the form
\begin{align}
\mathrm{d}s^2 = g_{\mu\nu} \mathrm{d}x^\mu \mathrm{d}x^\nu
= \frac{1}{A} \Big(-Q \mathrm{d}t^2
+ \frac{\mathrm{d}r^2}{Q}
+ r^2 e^{\Phi(x^i)} \delta_{ij} \mathrm{d}x^i \mathrm{d}x^j\Big),
\end{align}
in which
\begin{align*}
& A(r,x^i)
= (1 + \alpha r Z)^2, \\
& e^{\Phi(x^i)}
= \frac{(1 +  \alpha r_h Z)(1 - Z^2)}{(1+ \alpha r_h)^2
\delta_{ij} x^i x^j}, \qquad (i=1,2)
\end{align*}
where
\[
Z(x^i) = \cos \big(\theta (x^i)\big)
\]
is an implicit function of $x^i$. For calculation purposes, it is desirable to
rewrite the expression $\delta_{ij} x^i x^j$ in terms if the implicit function
$Z(x^i)$,
\begin{align}
w^2= \delta_{ij} x^i x^j
= \Bigg( \frac{1-Z}{1+Z}
\Big(\frac{(1 + \alpha r_h Z)^2}{1 - Z^2}\Big)^{\alpha r_h}
\Bigg)^{\frac{1}{1 - \alpha r_h}}     \label{Z}
\end{align}
because of the simple relation
\[
\partial_i Z = - (1+ \alpha r_h) e^\Phi x_i,
\]
which can be verified directly. This simple relation will
be very facilitating in the following calculations.

\section{Constraints on initial hypersurface}

We start by foliating the C-metric background by
three-dimensional timelike hypersurfaces defined by $r=const$.
The bulk line element could be expressed as
\[
\mathrm{d}s^2 = \frac{\mathrm{d}r^2}{AQ}
+ \gamma_{ab}\mathrm{d}x^a\mathrm{d}x^b
= \frac{\mathrm{d}r^2}{AQ}
+ \frac{1}{A} \Big(-Q \mathrm{d}t^2
+ r^2 e^{\Phi} \delta_{ij}
\mathrm{d}x^i \mathrm{d} x^j\Big),
\]
where $x^a=(t,x^i)$, and $\gamma_{ab}$ with $r$ taken to be constant is the
induced metric on each hypersurface. This foliation will enable us to consider
the ``initial value formulation'' of the vacuum Einstein's equation, taking
$r$ to be the evolution parameter\footnote{Regardless of the fact that $r$ is actually a
spacelike coordinate.}. The
appropriate initial data could be chosen as a hypersurface
$\Sigma_c$ located at $r=r_c$, together with it's first and
second fundamental forms. The first fundamental form is the
projection tensor
\[
\mathcal{P}_{\mu\nu} = g_{\mu\nu} - n_\mu n_\nu,
\]
where
\[
n_\mu = \frac{1}{\sqrt{AQ}}(\mathrm{d}r)_\mu
= \Big(0,\frac{1}{\sqrt{AQ}},0,0\Big),
\]
is the unit normal covector. This first fundamental form is
closely related to the induced metric, in our coordinate,
$ \mathcal{P}_{ab} = \gamma_{ab} $,
$ \mathcal{P}_{\mu r} = \mathcal{P}_{r \mu} = 0 $.
The second fundamental form is the extrinsic curvature of the
hypersurface
\begin{align}
K_{\mu\nu} = \frac{1}{2} \mathscr{L}_n \mathcal{P}_{\mu\nu},
\label{sec}
\end{align}
also in our choice of the coordinate system we have
$ K_{\mu r} = K_{r \mu} = 0 $.
After foliating the background manifold, tensors of type
$(0,2)$ could be decomposed into the following form
\[
S_{\mu\nu} = S_{\rho\sigma} \mathcal{P}^\rho{}_\mu
\mathcal{P}^\sigma{}_\nu + n_\mu (S_{\rho\sigma}
\mathcal{P}^\rho{}_\nu n^\sigma)
+ n_\nu (S_{\rho\sigma} \mathcal{P}^\rho{}_\mu n^\sigma)
+ n_\mu n_\nu (S_{\rho\sigma} n^\rho n^\sigma),
\]
so the equivalent form of the vacuum Einstein's equation is
\begin{equation}\label{CEID}
\begin{split}
& G_{\mu\nu} n^\mu n^\nu = 0,               \\
& G_{\mu\nu} \mathcal{P}^\mu{}_\rho n^\nu = 0,    \\
& G_{\mu\nu} \mathcal{P}^\mu{}_\rho \mathcal{P}^\nu{}_\sigma=0,
\end{split}
\end{equation}
the first two lines in \eqref{CEID} are the constraint equations of
the initial data $(\mathcal{P}_{\mu\nu},K_{\mu\nu})$ on the initial
hypersurface, and the third line is the evolution equation. According
to the Gauss-Codazzi equations the constraint equations could be cast
in the following form
\begin{subequations}\label{CE1}
\begin{align}
&\hat{R} + K^{ab} K_{ab} - K^2 = 0,  \label{hamiltonin c}\\
&D_a (K^a{}_b - \gamma^a{}_b K) = 0, \label{momentum c}
\end{align}
\end{subequations}
\noindent where $\hat{R}$ is the Ricci scalar of $\Sigma_c$, $D_a$ is the
covariant derivative compatible with $\gamma_{ab}$.
The equations \eqref{CE1} are often referred to as the Hamiltonion
and momentum constraints, respectively. Equivalently we could choose
$(\gamma_{ab},T_{ab})$ as the initial data, here
$T_{ab} = \gamma_{ab} K - K_{ab}$ is the the Brown-York
stress energy tensor, with the unit $8\pi G = 1$. Then the constraint equations
can be reformulated as
\begin{subequations} \label{CE2}
\begin{align}
& \mathscr{H} = \hat{R} + T^a{}_b T^b{}_a
- \frac{T^2}{2} = 0,   \label{hamc}     \\
& \mathscr{P}_b = D_a T^a{}_b = 0.       \label{momc}
\end{align}
\end{subequations}

Next we turn to the evolution equations $G_{\mu\nu}
\mathcal{P}^\mu{}_\rho \mathcal{P}^\nu{}_\sigma = 0$.
Rather than list their concrete forms expressed in terms of
$(\gamma_{ab},T_{ab})$, let us directly come to the
following conclusion,
i.e. {\it if we perturb the initial data which generate
the background spacetime, and demand that it evolves
no singularity in the bulk, then the geometry of the
perturbed spacetime should be of Petrov type I}. So there
are additional constraints of the initial data
\begin{align}
\mathscr{C}_{ij} =
l^\mu(m_i )^\nu l^\sigma(m_j)^\rho
C_{\mu\nu\rho\sigma} \big|_{\Sigma_c} = 0,
\label{PTI}
\end{align}
where $C_{\mu\nu\rho\sigma}$ is the bulk Weyl tensor, and
\begin{equation}
\begin{split}
&l^\mu = \frac{1}{\sqrt{2}}
\Big(\frac{\sqrt A_c}{\sqrt Q_c}(\partial_t)^\mu - n^\mu\Big), \\
&k^\mu = - \frac{1}{\sqrt{2}}
\Big(\frac{\sqrt A_c}{\sqrt Q_c}(\partial_t)^\mu + n^\mu\Big), \\
&(m_i)^\mu = r^{-1} e^{-\frac{1}{2}\Phi}
\sqrt{A_c} (\partial_i)^\mu,
\end{split}          \label{NPB}
\end{equation}
are a set of Newman-Penrose basis vector fields
located at the initial hypersurface, here $Q_c = Q(r_c)$,
$A_c = A(r_c,x^i)$. Inserting \eqref{NPB} into \eqref{PTI}, we get
\begin{align}
\frac{A_c}{Q_c} C_{titj}
+ \frac{\sqrt A_c}{\sqrt Q_c} C_{tij(n)}
+ \frac{\sqrt A_c}{\sqrt Q_c}C_{tji(n)}
+ C_{i(n)j(n)}=0, \label{pjbc}
\end{align}
and expressing these projections of the bulk Weyl tensor
by $(\gamma_{ab},T_{ab})$, the additional constraint
equations will finally become
\begin{equation}
\begin{split}
\mathscr{C}_{ij} =& 2\frac{Q_c}{A_c} T^t{}_i T^t{}_j
+ \frac{T^2}{4} \gamma_{ij}
- (T^t{}_t - 2 \frac{\sqrt{A_c}}{\sqrt{Q_c}} D_t)
\bigg(\frac{T}{2} \gamma_{ij} - T_{ij}\bigg)
\\ -& 2 \frac{\sqrt{Q_c}}{\sqrt{A_c}}
D_{(i} T^t{}_{j)} - T_{ik} T^k{}_j
- \hat{R}^t{}_{itj} - \hat{R}_{ij}=0, \label{ptc1}
\end{split}
\end{equation}
where $\hat{R}^a{}_{bcd}$, $\hat{R}_{ab}$ represent the Riemann
and Ricci tensors  of $\Sigma_c$. Up till now we have derived
all the constraint equations in our initial value formulation,
and in the following sections we will see that, on highly
accelerated hypersurface these equations give rise to the
Navier-Stokes equation.

\section{Non-relativistic hydrodynamic expansion
and constrained fluctuations}

For the background initial data, $\gamma_{ab}^{(B)}$ can be
read directly from the line element of initial hypersurface
$\Sigma_c$, so we can obtain the the background Brown-York
tensor $T_{ab}^{(B)}$, and the only nonzero components are
\begin{equation}
T^t{}_t^{(B)} = \frac{2\sqrt{Q_c}}{r_c},
\qquad T^i{}_j^{(B)} = \Big(\frac{\sqrt{Q_c}}{r_c}
- \alpha Z \sqrt{Q_c}
+ \frac{Q'_c \sqrt{A_c}}{2\sqrt{Q_c}}\Big)\delta^i{}_j,              \label{tb}
\end{equation}
here we have used the notations $Q'_c = Q'(r)|_{r=r_c}$,
$A'_c(x^i) = \partial_r A(r,x^i)|_{r=r_c}$ for short,
and in the rest of this paper the notations $Q'_h, Q''_h,
A'_h, A''_h$ will be similar to $Q'_c, Q''_c, A'_c, A''_c$ with
$r_c$ replaced by $r_h$, which represents the radial position of the black
hole event horizon. Before imposing perturbation to the
background metric, we first consider the non-relativistic
limit which will be essential when constructing the
non-relativistic hydrodynamics. This could be achieved
mathematically by rescaling the $t$ coordinate,
$t = \dfrac{\tau}{\lambda \sqrt{Q_c}}$, then the
back ground metric becomes
\begin{align}
\gamma_{ab}\mathrm{d}x^a\mathrm{d}x^b
&= \frac{1}{A_c} \Big(- \frac{1}{\lambda^2} (\mathrm{d} \tau)^2
+ r_c^2 e^{\Phi} \delta_{ij}
\mathrm{d}x^i \mathrm{d} x^j\Big), \label{dsd}
\end{align}
the reciprocal of the rescaling parameter $\lambda$ can viewed as
the speed of light and $\lambda \rightarrow 0$
corresponds to the non-relativistic limit. By explicit
calculations we find that the Brown-York tensor $T^a{}_b^{(B)}$
and the constraint equations \eqref{CE2} are kept invariant
under this rescaling, whilst some coefficients in the
additional constraint \eqref{ptc1} is changed:
\begin{equation}
\begin{split}
\mathscr{C}_{ij}
=& \frac{2}{\lambda^2}\frac{1}{A_c} T^\tau{}_i T^\tau{}_j
+ \frac{T^2}{4} \gamma_{ij}
- (T^\tau{}_\tau - 2 \lambda \sqrt{A_c} D_\tau)
\bigg(\frac{T}{2} \gamma_{ij} - T_{ij}\bigg)
\\ -& \frac{2}{\lambda} \frac{1}{\sqrt{A_c}}
D_{(i} T^\tau{}_{j)} - T_{ik} T^k{}_j
- \hat{R}^\tau{}_{i\tau j} - \hat{R}_{ij}=0. \label{ptc2}
\end{split}
\end{equation}
Then we take into account the hydrodynamic limit. As is proven
in\cite{strominger1}, there is a mathematical equivalence between the
near horizon limit and the hydrodynamic limit. So let us
take a particular initial hypersurface which is close to the
black hole horizon at $r = r_h$. This can be realized by introducing a
small parameter $\epsilon$ via $r_c - r_h = \epsilon^2$. We
would like to link the two small parameters $\epsilon$ and $\lambda$ via
$\epsilon=\chi \lambda$, where
the constant $\chi$ is employed to balance the  dimensionality.
This identification makes the non-relativistic limit and the
hydrodynamic limit occur simultaneously by taking
$\lambda \rightarrow 0$. In this limit the background Brown-York
tenser  $T^a{}_b^{(B)}$ could be expanded as
\begin{equation}
\begin{split}
&T^\tau{}_\tau^{(B)} = 2 \chi \lambda \frac{\sqrt{Q'_h}}{r_h}
+ \cdots,
\qquad T^t{}_i^{(B)} = 0,        \\
&T^i{}_j^{(B)} = \bigg( \frac{1}{2}
\frac{\sqrt{A_h Q'_h}}{\chi \lambda}
+ \chi \lambda \Big(\frac{\sqrt{Q'_h}}{r_h}
- \frac{1}{2} \alpha Z \sqrt{Q'_h}
+ \frac{1}{2} \frac{ Q''_h \sqrt{A_h}}{\sqrt{Q'_h}}\Big)
\bigg)\delta^i_{\ j} + \cdots.
\label{BBY}
\end{split}
\end{equation}

Now let us consider the perturbation theory in the initial
value formulation, the most general equations for the fluctuations
around the background initial data can be very complicated,
so we restrict ourselves to the following non-relativistic
hydrodynamic $\lambda$-expansion.
\begin{subequations}
\begin{align}
\gamma_{ab} &= \gamma_{ab}^{(B)}+\sum_{n=1}^\infty
\gamma_{ab}^{(n)}\lambda^n, \label{gamfluc}   \\
T^a{}_b &= T^a{}_b^{(B)} + \sum^\infty_{n=1} \lambda^n
T^a{}_b^{(n)},          \label{seby}
\end{align}
\end{subequations}
where $(\gamma_{ab}^{(n)},T^a{}_b^{(n)})$ represent the
fluctuation modes. In the above double expansion the induced
geometry of the initial hypersurface is perturbed, by explicit
calculations we can list all the expansions of the Christoffel
symbol, the Riemann tensor $\hat R_{abcd}$ and of the Ricci
tensor $\hat R_{ab}$, subjecting to the perturbed metric
$\gamma_{ab}$, but for what we are considering, the only necessary expression is $\hat{R}^\tau{}_{i\tau j}
+ \hat{R}_{ij}$
\begin{align}
\hat{R}^\tau{}_{i\tau j}
+ \hat{R}_{ij}
= \Big(\frac{Q_h}{r_h^2} - \frac{2 \alpha Z Q_h}{r_h}
+ \frac{Q'_h \sqrt{A_h}}{r_h}\Big) \gamma_{ij}^{(B)}
+ \cdots.
\label{backR}
\end{align}
with all of this materials, next we will
derive the constraint equations for the fluctuations.

\textbf{(a) Perturbed Hamiltonian constraint}

The middle term in the Hamiltonian constraint \eqref{hamc} can be further
decomposed according to the spacial slicing, i.e.
\begin{equation}
\mathscr{H} = \hat{R} + 2 T^\tau{}_i T^i{}_\tau
+ T^j{}_i T^i{}_j - \frac{T^2}{2} = 0,
\end{equation}
according to \eqref{seby} and \eqref{backR}, we can get
\begin{equation}
\begin{split}
&\hat{R} = 2 \frac{Q'_h \sqrt{A_h}}{r_h}
+ \mathcal{O}(\lambda^1) ,      \\
& T^\tau{}_i T^i{}_\tau
= -\frac{\gamma^{ij(0)}}{A_h}
T^\tau{}_i^{(1)} T^\tau{}_j^{(1)}
+ \mathcal{O}(\lambda^1) ,  \\
& \frac{T^2}{2} = \frac{1}{2}
\frac{Q'_h A_h}{(\chi \lambda)^2}
+ 4 \frac{Q'_h \sqrt{A_h}}{r_h}
- \alpha Z Q'_h \sqrt{A_h} +  Q''_h A_h
+ \frac{\sqrt{Q'_h A_h}}{\chi} T^{(1)}
+ \mathcal{O}(\lambda^1),
\\& T^i{}_j T^j{}_i = \frac{1}{2}
\frac{Q'_h A_h}{(\chi \lambda)^2}
+ 2 \frac{Q'_h \sqrt{A_h}}{r_h}
- \alpha Z Q'_h \sqrt{A_h} + Q''_h A_h
+ \frac{\sqrt{Q'_h A_h}}{\chi} T^i{}_i^{(1)}
+ \mathcal{O}(\lambda^1),
\end{split}
\end{equation}
where $\gamma^{ik(0)} = A_h r_h^{-2} e^{-\Phi}\delta^{ik}$,
so, at the first nontrivial order of the perturbed
Hamiltonion constraint, we have
\begin{align}
\mathscr{H}^{(0)} = 0 \quad\Longrightarrow\quad
T^\tau{}_\tau^{(1)} = -2 \frac{\chi}{\sqrt{Q'_h A_h^3}}
\gamma^{ij(0)} T^\tau{}_i^{(1)} T^\tau{}_j^{(1)}.  \label{hamc1}
\end{align}

\textbf{(b) expansions for momentum and additional
constraint}

For the general case
when both induced metric and and Brown-York tensor receive perturbation the covariant form of the momentum
and additional constraint will not be strictly expanded as a
power series in $\lambda$, so we need to express these constraint equations
by ordinary derivative.
Firstly inserting \eqref{seby} into \eqref{momc}
\[
D_a T^a{}_b = D_a T^a{}_b^{(B)} +
\lambda D_a T^a{}_b^{(1)} + \cdots,
\]
\begin{subnumcases}{\Rightarrow}
\tau\text{-component}:& $D_a T^a{}_\tau = - \dfrac{1}{\lambda}
\dfrac{\gamma^{ij(0)}}{A_h}D_i T^t{}_j^{(1)} + \cdots$,    \\
i\text{-component}:&
$
D_a T^a{}_i = \lambda \Big(D_\tau T^\tau{}_i^{(1)}
+ D_j T^j{}_i^{(1)}
- \chi \dfrac{\alpha x_i \sqrt{Q'_h}}{2\sqrt{A_h}}
e^\Phi\Big) + \cdots.
$                 \label{mccd}
\end{subnumcases}
so at the first nontrivial order of the momentum constraints, we get
\begin{subequations}\label{momcp2}
\begin{align}
&\mathscr{P}_\tau^{(-1)} =0 : \,
\gamma^{ij(0)}(\partial_i
+ 3 \frac{(1+ \alpha r_h)\alpha r_h x_i}{\sqrt{A_h}} e^\Phi)
T^{\tau(1)}_{\ j} = 0,   \label{momcpt2}\\
&\mathscr{P}_i^{(1)} = 0 : \,
\partial_\tau T^\tau{}_i^{(1)}
+\Big(\partial_j + \partial_j \Phi
+ 3 \frac{(1+ \alpha r_h)\alpha r_h e^\Phi}{\sqrt{A_h}}
x_j\Big) T^j{}_i^{(1)}  \nonumber\\
&\qquad\qquad\qquad\quad
-\frac{1}{2} T^j{}_j^{(1)} \partial_i \Phi
- \frac{(1+\alpha r_h)\alpha r_h e^\Phi}{\sqrt{A_h}} x_i T^{(1)}
- \chi \frac{(1+\alpha r_h)\alpha x_i \sqrt{Q'_h}}{2\sqrt{A_h}}
e^\Phi = 0,      \label{momcpi2}
\end{align}
\end{subequations}
due to \eqref{seby} and \eqref{backR}
\begin{align*}
& \frac{T^2}{4} = \frac{1}{4}
\frac{Q'_h A_h}{(\chi \lambda)^2}
+ 2 \frac{Q'_h \sqrt{A_h}}{r_h}
- \frac{1}{2} \alpha Z Q'_h \sqrt{A_h}
+ \frac{1}{2} Q''_h A_h
+ \frac{\sqrt{Q'_h A_h}}{\chi} \frac{T^{(1)}}{2}
+ \mathcal{O}(\lambda^1),
\\& T^i{}_k T^k{}_j = \Big(\frac{1}{4}
\frac{Q'_h A_h}{(\chi \lambda)^2}
+ \frac{Q'_h \sqrt{A_h}}{r_h}
- \frac{1}{2} \alpha Z Q'_h \sqrt{A_h}
+ \frac{1}{2} Q''_h A_h\Big) \delta^i{}_j
+ \frac{\sqrt{Q'_h A_h}}{\chi} T^i{}_j^{(1)}
+ \mathcal{O}(\lambda^1),
\\& \frac{T}{2} \delta^i{}_j- T^i{}_j
= \lambda \Big(\frac{T^{(1)}}{2} \delta^i{}_j
- T^i{}_j^{(1)} + \chi \frac{\sqrt{Q'_h}}{r_h}
\Big) + \mathcal{O}(\lambda^1),
\end{align*}
so, the first nontrivial order of the additional constraint
\eqref{ptc2} will be $\mathcal{O}(\lambda^0)$, and we have
\begin{align}
\mathscr{C}^i{}_j^{(0)} = 0 : \, T^{i(1)}_{\ j}
&= \frac{\chi}{\sqrt{Q'_h A_h}} \cdot 2 \gamma^{ik(0)}
\Big(\frac{1}{A_h} T^\tau{}_k^{(1)} T^\tau{}_j^{(1)}
- \frac{1}{\sqrt{A_h}} \zeta_{kj}\Big)
+ \frac{T^{(1)}}{2} \delta^{i}_{\ j},
\label{addcp2}
\end{align}
where we have used the following short-hand notation
\begin{align}
\zeta_{kj}
=& \partial_{(k} T^\tau{}_{j)}^{(1)}
- \partial_{(k} \Phi T^\tau{}_{j)}^{(1)}
+ \frac{1}{2} \delta_{kj} \delta^{lm}
\partial_l\Phi T^\tau{}_m^{(1)}      \nonumber\\
&\quad- \frac{(1+ \alpha r_h)\alpha r_h}{\sqrt{A_h}} e^\Phi
(x_{(k} T^\tau{}_{j)}^{(1)} - \delta_{kj} \delta^{lm}
x_l T^\tau{}_m^{(1)}).      \label{zeta}
\end{align}
Interestingly, from \eqref{momcpt2} and \eqref{addcp2}
we find that
\begin{align}
\mathscr{C}^j{}_j^{(0)} = 0 \Rightarrow
T^\tau{}_\tau^{(1)} = -2 \frac{\chi}{\sqrt{Q'_h}}
\frac{\gamma^{ij(0)}}{A_h}
\Big(\frac{1}{\sqrt{A_h}}
T^\tau{}_i^{(1)} T^\tau{}_j^{(1)}
+ 3\frac{(1+ \alpha r_h)\alpha r_h e^\Phi}{\sqrt{A_h}}
x_i T^\tau{}_j^{(1)}\Big),   \label{cjj}
\end{align}
then compare \eqref{cjj} with \eqref{hamc1} we will come
to the following relation
\begin{align}
\alpha \cdot \delta^{ij} x_i T^\tau{}_j^{(1)} = 0,  \label{vortex}
\end{align}
from eqs.\eqref{hamc1}, \eqref{momcp2} and \eqref{addcp2} we could
also establish the continuity and Navier-Stokes equation,
but these equations are
not expressed in a covariant form, so we would rather interpret
the corresponded system as fluid lives in Newtonian spacetime, further
the equation \eqref{vortex} implies that there is a vortex
in the fluid system.

\section{Vortex fluid in flat space}

In this section we will study the hydrodynamic equations
constructed from the constraints of the fluctuations derived in
the last section, and turn our attention only to the strictly expanded constraint
equations with $\alpha \neq 0$, because the case $\alpha=0$ corresponds to
fluctuations around Schwarzschild black hole solution which is somewhat well
understood.  Since $\alpha \neq 0$, eq.\eqref{vortex} yields
\begin{align}
\delta^{ij} x_i T^\tau{}_j^{(1)} = 0, \label{crtc}
\end{align}
which turns out to be a critical new condition for this case.
First of all, the condition \eqref{crtc} implies
that eq.\eqref{momcpt2} can be rearranged into
\begin{align}
\partial^j  T^\tau{}_j^{(1)} =0. \label{ctn}
\end{align}
To analyze the ``spatial'' components of the perturbed momentum
constraint, we need to insert \eqref{addcp2} and \eqref{cjj}
into \eqref{momcpi2}. The computation is rather complicated, but
at the first nontrivial order the equation can be simplified
into the following form
\begin{align}
& \partial_\tau T^\tau{}_i^{(1)}
+ \frac{1}{2} \partial_i T^{(1)}
+ \frac{\delta^{jk}}{r_h^2 e^\Phi}
\bigg[\frac{2}{\sqrt{A_h}}
T^\tau{}_k^{(1)} \partial_j T^\tau{}_i^{(1)}
- \partial_j \partial_k T^\tau{}_i^{(1)}
- \frac{2}{\sqrt{A_h}}
T^\tau{}_j^{(1)} T^\tau{}_k^{(1)}
\partial_i \Phi               \nonumber\\
& + \partial_j \Phi (\partial_k T^\tau{}_i^{(1)}
-  \partial_i T^\tau{}_k^{(1)})
+ T^\tau{}_i^{(1)} \partial_j \partial_k\Phi
- \frac{(1+ \alpha r_h)\alpha r_h e^\Phi}{\sqrt{A_h}}
\Big(2x_j \partial_i T^\tau{}_k^{(1)}
+ 2x_j \partial_k T^\tau{}_i^{(1)}    \nonumber\\
&-2 (\delta_{jk} + 2 x_j \partial_k \Phi
+ 2\frac{(1+ \alpha r_h)\alpha r_h e^\Phi}{\sqrt{A_h}}
x_j x_k) T^\tau{}_i^{(1)}\Big)\bigg]
+ \frac{(1+\alpha r_h)\alpha r_h e^\Phi}{2\sqrt{A_h}} \Big(T^{(1)}
- \frac{Q'_h}{r_h}\Big) x_i = 0. \label{NSeq2}
\end{align}

Next we are going to
interpret eqs. \eqref{ctn} and \eqref{NSeq2} as the continuity
and the  Cauchy momentum equations in flat Newtonion spacetime,
so in what follows, all the indices will be raised and lowered
by $\delta^{ij}$ and its inverse $\delta_{ij}$. As the last step
we would like to introduce the following
``holographic dictionary''
\[
\quad T^\tau{}_i^{(1)} = \rho  v_i  ,\quad
\quad T^{(1)} = 2 p,
\]
where $\rho, v_i, p$ represent the density distribution,
velocity field and the pressure of the fluid system. Under this dictionary eq.\eqref{ctn} becomes the continuity equation
\begin{align}
\partial^j (\rho  v_j) = 0,  \label{Cont}
\end{align}
with the density distribution $\rho = \frac{\sqrt {A_h}}{2} r_h e^\Phi$,
and eq.\eqref{NSeq2} becomes the Navier-Stokes equation for the viscous fluid
\begin{align}
\rho (\partial_\tau v_i + v^j \partial_j v_i)
+ \partial_i p - \eta \partial^j \partial_j v_i = f_i, \label{NV}
\end{align}
with the dynamic viscosity
$\eta=\frac{\sqrt{A_h}}{2}$. For C-metric black hole in isotropic
coordinate, if the trivial infinite factor in the black area
is taken to be the volume parallel to the event horizon, then the
entropy density should be
$s=\frac{1}{4G\sqrt{A_h}(1+\alpha r_h)^2}$,
therefore the viscosity to entropy ratio is
\[
\frac{\eta}{s} = \frac{1}{4\pi}(1+\alpha r_h Z)^2(1+\alpha r_h)^2,
\]
we see that for accelerated black hole the shear viscosity to entropy density ratio is affected by the black hole acceleration
, when $\alpha=0$, this ratio will be $\frac{1}{4\pi}$. Then For the right hand side, $f_i$ represents the
body force density
\begin{align}
f_i =&  F(Z)\big(2 \mu x^j \partial_{[j} v_{i]}
- 2\mu  v_i + \rho v^2 x_i\big)  \nonumber  \\
&-  \frac{(1+2\alpha m) \alpha}{\nu r_h}
\Big[ \Big(2p - \frac{Q'_h}{r_h}\Big)\frac{x_i}{2\mu} +
\Big(4 +2 F(Z) w^2
+3 \frac{(1+2\alpha m)\alpha}{2 \mu \nu r_h} w^2 \Big)v_i\Big] , \label{flatforce}
\end{align}
in the last force term, we have used
the short-hand notation
\[
F(Z)=\frac{(Z-1)(3\alpha r_h Z +3\alpha r_h +2)}{w^2(1+\alpha r_h)},
\]
it is clear that the second line in eq.\eqref{flatforce} is a pure
inertia force because of the overall factor $\alpha$. As for the first line
in eq.\eqref{flatforce}, we could recognize the first term as a kind of Coriolis
force and the second term as a linear resistance force, so only the last term on
the first line remains a mystery. Nevertheless one can recognize that this
last force term is proportional to the kinetic energy density of the fluid
component, which also sounds good. Anyway we should not forget the role of
\eqref{crtc}, which is now rewritten in terms of the holographic dictionary as
\begin{align}
x^i v_i=0,      \label{vc}
\end{align}
i.e. the velocity field of the fluid is always perpendicular to its spacial
displacement. This is nothing but a vortex condition. In other words, the fluid
we constructed on the flat Newtonian spacetime is precisely a holographic vortex, when $w \rightarrow \infty$,
which correspond to the region near the conical singularity,
the asymptotic behaviour of the velocity field is
\[
v \sim w^{\frac{2}{1+\alpha r_h}}
\sin\Big(\frac{2\sqrt{\alpha r_h}}{1+\alpha r_h} \log w\Big),
\]
it can be seen from that the velocity
field does not vanish in the far zone. This effect can be interpreted by the
presence of the Coriolis-like force term in the fluid equation.

\section{Conclusion}

Unlike the ordinary static black holes with maximally symmetric horizons, the
C-metric black hole represents two black holes under constant proper acceleration.
The acceleration of the black hole squeezes the horizon surface, leaving less symmetries
than the non-accelerating black holes. Our construction reveals that Gravity/Fluid
correspondence can be realized in terms of Petrov I boundary condition even for black holes with
less symmetries than the usual static black holes with maximally symmetric horizons.

To be more concrete, we have realized a fluid system
from the vacuum C-metric black hole solution, which lives on a flat Newtonian
spacetime, and possess non-constant but stationary density distributions
and kinematic viscosities, so it is compressible
viscous fluids subject to extra body forces.
Compared with previous studies on Gravity/Fluid correspondences, the present work
differs in two major aspects. The first difference of our present work
lies in that the extra body forces arising from the C-metric black
hole case can have more appropriate physical interpretations. For the flat space fluid 
system, the extra force are consisted of
an inertial force term, a Coriolis-like term, a linear resistance term and a
term proportional to the kinetic energy density of the fluid. It is remarkable that
the combination of all these complicated force terms gives rise to a pure vortex
behavior for this case. The second difference
lies in that the shear viscosity to entropy density ratio of the corresponded system 
is subjected to the black hole acceleration. Besides, just as those anisotropic theories 
which violate the KSS bound
\cite{Rebhan:2011vd,Critelli:2014kra,Ge:2014aza}, for the accelerated
black hole with less symmetries, the ratio is also lower than the bound in the region around the conical singularity.

Before ending, let us stress that the Gravity (with curved horizon)/Flat space fluid
correspondence realized in \cite{HWZ,Hao:2015zxa} and the present work seems to rely on
the conformal flatness of the horizon surface of the background geometry. However,
going through the details of the construction, it is evident that such correspondence only
requires the existence of a map from the near horizon hypersurface to the flat space,
be it conformal or not. Therefore, it is
tempting to consider other cases with more complicated, less symmetric black hole backgrounds.
Doing so one might be able to get more general fluid systems with less constraints on the
density distributions and/or kinematic viscosities. For this purpose, the black ring geometry
in 5 dimensions may be a good choice as background geometry. We leave the study of such
backgrounds to later works.

\section*{Acknowledgement}

This work is supported by the National Natural Science Foundation of China under the grant
No. 11575088 and No. 11605137.

\providecommand{\href}[2]{#2}\begingroup
\footnotesize\itemsep=0pt
\providecommand{\eprint}[2][]{\href{http://arxiv.org/abs/#2}{arXiv:#2}}


\end{document}